\documentstyle[12pt,twoside,fleqn,espcrc1]{article}
\def\apj #1 #2 #3 {ApJ. #1 (#2) #3}                     
\def\apjl #1 #2 #3 {ApJ. #1 (#2) L#3}   
\def\apjs #1 #2 #3 {ApJS #1 (#2) #3}   
\def\aap  #1 #2 #3 {#1, A\&A, #2, #3}
\def\mnras #1 #2 #3 {#1, MNRAS, #2, #3}         
\def\pra #1 #2 #3 {Phys.~Rev.~A.~#1 (#2) #3}
\def\prb #1 #2 #3 {Phys.~Rev.~B.~#1 (#2) #3}   
\def\prc #1 #2 #3 {Phys.~Rev.~C.~#1 (#2) #3}
\def\prd #1 #2 #3 {Phys.~Rev.~D.~#1 (#2) #3}
\def\pre #1 #2 #3 {Phys.~Rev.~E.~#1 (#2) #3}   
\def\prl #1 #2 #3 {Phys.~Rev.~Lett. #1 (#2) #3}        
\def\plb #1 #2 #3 {Phys.~Lett.~B.~#1 (#2) #3}
\def\science #1 #2 #3 {Science.~#1 (#2) #3}   
\def\nature #1 #2 #3 {Nature.~#1 (#2) #3}   
\def\nphysa #1 #2 #3 {Nucl.~Phys.~A~#1 (#2) #3}       
\def\nphysb #1 #2 #3 {Nucl.~Phys.~B~#1 (#2) #3}       
\def\nphysbs #1 #2 #3 {Nucl.~Phys.~B.~Suppl. #1 (#2) #3}

\def\he#1{\hbox{${}^{#1}$He}}
\def\li#1{\hbox{${}^{#1}$Li}}

\def\yp{\hbox{$Y_{\rm p}$}}

\def\omegab{\hbox{$\Omega_b$}}

\def\mev{\mbox{~MeV}}

\def\la{\mathrel{\mathpalette\fun <}}
\def\ga{\mathrel{\mathpalette\fun >}}
\def\fun#1#2{\lower3.6pt\vbox{\baselineskip0pt\lineskip.9pt
  \ialign{$\mathsurround=0pt#1\hfil##\hfil$\crcr#2\crcr\sim\crcr}}}

\newcommand{\AmS}{{\protect\the\textfont2
  A\kern-.1667em\lower.5ex\hbox{M}\kern-.125emS}}

\title{Neutrinos and Big-Bang Nucleosynthesis}

\author{T. KAJINO
      \address{National Astronomical Observatory, Division 
       of Theoretical Astrophysics}
      \address{The Graduate University for Advanced Studies, 
       Department of Astronomical Science \\
       2-21-1 Osawa, Mitaka, Tokyo 
       181, Japan}
       \address{University of Tokyo, Department of Astronomy \\
       2-11-16 Yayoi, Bunkyo-ku, Tokyo 113, Japan}        
       and 
        M. ORITO$^{\rm a}$}

\begin{document}
\maketitle
\input epsf
\begin{abstract}
Observations of clusters and super clusters of galaxies have indicated 
that the 
Universe is more dominated by baryons than ever estimated in the homogeneous 
cosmological model for primordial nucleosynthesis.  Recent detections of 
possibly low deuterium abundance in Lyman-$\alpha$ clouds along 
the line of sight to high red-shift quasars have raised another potential 
difficulty that \he4 is overproduced in any cosmological models which 
satisfy the low deuterium abundance constraint.
We show that the inhomogeneous cosmological model with degenerate 
electron-neutrino can resolve these two difficulties.
\end{abstract}

\section{Introduction}

One of the cosmological impacts of primordial nucleosynthesis is on 
the universal baryon mass density $\rho_b $.  It is of 
significance to answer the question 
how much fraction of universal mass is made of ordinary matter baryons.
Homogeneous Big-Bang model for primordial 
nucleosynthesis~\cite{wagoner67,copi95}, assuming the standard model for light neutrino families, 
predicts small $\Omega_b$, $0.03 \le \Omega_b h_{50}^2 \le 0.06$, where 
$\Omega_b = \rho_b / \rho_c$, $\rho_c$ is the critical density which 
marginally closes the Universe, and $h_{50}$ is the Hubble constant 
$H_0$ divided by $50\,km/s/Mpc$. However, recent observations of rich 
clusters and super clusters of galaxies have indicated much larger baryon 
fraction, 0.1 $\le \Omega_b h_{50}^{3/2} \le$ 
0.3~\cite{white93,bahcall95}.

Inhomogeneous Big-Bang model~\cite{appligate85}--\cite{mathews96}, which allows inhomogeneous 
baryon density distribution due to various physical processes in the early 
Universe, has been proposed in order to resolve this discrepancy. In this 
model difference in diffusion effects between neutrons and charged nuclei 
plays an important role in fluctuating density distribution to 
suppress overproduction of \he4, and resultant $\Omega_b$ is relaxed to 
$\Omega_b \sim 0.1$~\cite{orito97}. 

However, another potential difficulty has been imposed by recent observations 
of deuterium absorption line in Lyman-$\alpha$ clouds along the line of sight 
to high red-shift quasars~\cite{rugers96,tytler96}. Several 
detections~\cite{tytler96} among them provide too small 
deuterium abundance to accept concordant $\Omega_b$ which satisfies 
the abundance constraints on the other light elements \he3, \he4 and \li7.
The observed deuterium abundance still scatters largely by one 
order of magnitude depending on different Lyman-$\alpha$ systems, 
and there are still many error sources unclear in the analysis of 
abundance determination. However, if these detections are real, 
the abundance found there is presumed to constrain most strongly 
the primordial abundance because these clouds are the primitive 
gas which still resides in the epoch of galaxy formation 
and has not been processed very much in its evolutionary 
history. 

It is the purpose of this paper to propose that the inhomogeneous 
cosmological model with degenerate electron-neutrino can resolve these 
two difficulties simultaneously within the framework of the standard model 
for neutrino. In the next section we first discuss neutrino properties 
in the early Universe which can  affect 
strongly the primordial nucleosynthesis. 
We then present the results of primordial 
nucleosynthesis calculated in both homogeneous and inhomogeneous cosmological 
models in sect. 3, and the $\Omega_b$ problem and the problem of 
overproduction of \he4 
are discussed in details. Finally, in sect. 4, we summarize this paper. 

\section{Neutrino in the Early Universe}

Radiation and relativistic particles played more important role than ordinary 
matter in the evolution of early hot Big-Bang Universe. Since relativistic 
neutrinos had energy density comparable to the densities due to photons and 
charged leptons, a small modification of neutrino properties can change the 
expansion rate of the Universe and a resultant slight shift of weak decoupling 
temperature affects strongly the primordial 
nucleosynthesis~\cite{weinberg72}.

Let us consider the \he4 synthesis in this section in order to make clear the 
role of neutrinos in primordial nucleosynthesis.  Although all produced 
elements are, in principle, influenced by the weak interactions, 
sensitivity of inferred $\Omega_b$ to primordial \he4 abundance 
is very critical to resolve the crises 
for the following two reasons~\cite{wagoner67}.  First, the neutron-to-proton 
ratio before nucleosynthesis is delicately determined by the weak 
interactions between nucleons and leptons, which 
controls sensitively the \he4 yield.
The second reason is that the \he4 abundance in HII regions is so 
accurately studied observationally that even one percent change in 
predicted abundance results in different \omegab value. 
This is to be compared 
with less accurate abundance constraints on the other elements like
D and \li7 whose primordial abundances are only known by order.  

\subsection{Neutron-to-Proton Ratio and Weak Decoupling}

The nucleosynthesis process can be studied quantitatively only by 
numerical calculation of solving a large number of non-linear rate 
equations because nuclear statistical equilibrium is not maintained 
down to $T \sim$ 100 keV ($\sim$ 10$^9$ K) at which explosive 
nucleosynthesis occurs. However, since the major nuclear reactions 
are mediated by the strong and electromagnetic interactions, except 
for beta decays of triton, $^7$Be, etc., total numbers of protons and 
neutrons are approximately conserved. This avoids complications in our 
qualitatively discussion of \he4 synthesis in this section. The only effect 
on the neutron-to-proton abundance ratio is the free decay of neutrons. 
Taking this into consideration, the ratio just before the onset of 
nucleosynthesis is given by
\begin{equation}
       {\rm{r}}(T) \equiv n_{\rm{n}} / n_{\rm{p}}= 
       {\rm{r}}(T_d) exp(-t/{\tau_n}),
	\label{eq:1}
\end{equation}
where $n_{\rm{n}}$ and $n_{\rm{p}}$ are the neutron and proton number 
densities, $\tau_n$ is the neutron mean life, and ${\rm{r}}(T_{d})$ is 
the ratio 
at decoupling temperature of weak interactions.

Weak decoupling is defined at the time when the collision time scale 
of weak interactions, 
$\tau_c  \sim  <n_\nu \sigma v>^{-1} = {G_F}^{-2} T^{-5}$, and 
the Hubble expansion 
time scale, $\tau_H  = R/\dot{R} = [1.66T^2\sqrt{g^*}/M_{Pl}]^{-1}$, 
balance with each other,  
where $G_F$ is Fermi coupling constant, $M_{Pl}$ Planck mass, $g^*$ 
the effective degrees of freedom of relativistic particles to be discussed 
later, and $T$ the universal temperature. Equating these two time scales, 
we obtain 
\begin{equation}
	T_d  \sim [1.66 \sqrt{g^*} /M_{Pl} / {G_F}^2]^{1/3}  \sim  
      O(1\mev).
	\label{eq:2}
\end{equation}
At earlier times $t \la 1$ sec and $T \ga T_d \sim$ 1 MeV, 
$\tau_c \la \tau_H$ holds and protons and neutrons are in chemical 
equilibrium by frequent weak interactions, n + $\nu_e$ 
$\leftrightarrow$ p + e$^-$, n + e$^+$ $\leftrightarrow$ p + $\bar{{\nu_e}}$, 
and n $\leftrightarrow$ p + e$^- + \bar{{\nu_e}}$. $r(T_{d})$ is thus given by 

\begin{equation}
	{\rm{r}}(T_{d}) = (m_{\rm{n}}/m_{\rm{p}})^{3/2} exp(- 
	\delta m/T_{d} - \mu_{\nu_e} / T_{d}),
	\label{eq:3}
\end{equation}
where $\delta m = m_{\rm{n}} - m_{\rm{p}} = 1.293$ MeV is the 
neutron-proton mass difference, and $\mu_{\nu_e}$ the chemical 
potential of electron-neutrino.  Shortly after the weak decoupling, 
electrons and positrons have pair annihilated to reheat the Universe for 
entropy conservation.  After this epoch the weak processes which 
interconvert neutrons and protons freeze out, and the abundance ratio 
r($T$) stays at a value close to r($T_d$).  

Once nucleosynthesis occurs explosively, nuclear 
reactions incorporate almost all available neutrons into $^4$He, 
which has the highest binding energy among all light nuclei.  
The Coulomb barrier for the \he4-induced reactions with other charged 
nuclei prevents them from destroying easily once it is created. 
Thus, the mass fraction of $^4$He is simply approximated by
\begin{equation}
	\yp  \sim  2n_{\rm{n}}/(n_{\rm{n}} + n_{\rm{p}}) = 
	2{\rm{r}}(T)/(1 + {\rm{r}}(T)),	
	\label{eq:4}
\end{equation}
at the risk of losing quantitative accuracy which should follow the 
full network 
calculation to be discussed in sect.~3.

\subsection{Partial Degenerate Neutrino}

An arbitrary conjecture which has been assumed in the standard 
cosmological model for primordial nucleosynthesis is that the net 
lepton number, 
$L_x = \{n_{\nu_x} - n_{\bar{\nu_x}} + n_x - n_{\bar x}\} /n_{\gamma}$ 
($x = e, \mu$ or $\tau$), is zero. However, 
any particle-physics and cosmological constraints do not rule out 
finite lepton asymmetry of the Universe within the framework of the 
standard model for quarks and leptons.  It is even 
natural to assume $L_x \neq 0$ cosmologically.

After the epoch of cosmic quark-hadron phase transition which occurred 
at $T \sim 100$ MeV, only abundant charged leptons were electrons and 
positrons.  For charge neutrality, however, $(n_{e^-} - n_{e^+})/n_{\gamma} 
= n_B/n_{\gamma} \le 10^{-8}$ for the $\Omega_b \le 1$ Universe models. 
Therefore, chemical potential of the electron must be negligibly small. 
Only neutrinos $\nu_e$, $\nu_\mu$ and $\nu_\tau$ and their antiparticles 
might have non-vanishing chemical potentials.  Since electron-lepton 
number and also muon- and tau-lepton numbers are believed to be conserved 
separately, the degeneracy parameters $\xi_{\nu_x}$ = $\mu_{\nu_x} / kT$ 
must be constant.

Let us consider partial degeneracy of the only electron-neutrino, 
$\left|{\xi_{\nu_e}}\right| \ll 1$, in order to see the effects on 
\he4 synthesis through eqs.~(1) to (4).  
We are not interested in degeneracy of the other 
neutrino species because these are insensitive to primordial 
nucleosynthesis. (see the discussion below).  
We also discuss neither complete neutrino degeneracy which has been studied in 
literature. 

There are two different effects of partial electron-neutrino 
degeneracy on nucleosynthesis.  The first effect arises from additional energy 
density due to degenerate neutrinos:
\begin{equation}
	(\rho_{\nu_e} +  \rho_{\bar{{\nu_e}}}) / \rho_{\gamma}  =  
(T_{\nu_e} / T_{\gamma})^4 \left[7/8 + 15/4{\pi}^2 {\xi_{\nu_e}}^2 
+ 15/8{\pi}^4 {\xi_{\nu_e}}^4\right].
	\label{eq:5}
\end{equation}
In the limit of vanishing chemical potentials for all relativistic particles, 
the effective degrees of freedom $g^*$ which appeared in 
eq.~(\ref{eq:2}) 
is given by $g^*  =  \sum_{b=Bosons} g_b  +  (7/8) \sum_{f=Fermions} g_f$.
The second and third terms of $r.h.s.$ in eq.~(5) are additional degrees of 
freedom for degenerate neutrinos, $\Delta g^*$.  Combining eqs.~(1) - (4) 
with this change from $g^*$ to $g^* + \Delta g^*$ and artificially 
suppressing the change in neutron-to-proton ratio due 
to $exp(- \mu_{\nu_e} / T$) 
in eq.~(3) , we can estimate the change in the mass fraction of \he4:
\begin{equation}
	\Delta \yp  \sim  1/6~\yp (1 - \yp /2) (\delta m / T_d)  
        \Delta g^* / g^*  > 0. \label{eq:delyp1}
     \label{eq:6}
\end{equation}
This is a second order effect of $\xi_{\nu_e}$, i.e. $\Delta g^* \sim 
{\xi_{\nu_e}}^2$.  

The second more important effect arises directly from the factor 
exp(- $\mu_{\nu_e} / T$) which we ignored artificially in the above 
discussion.  A simple calculation leads to the 
change in \yp
\begin{equation}
	\Delta \yp  \sim   - \yp (1 - \yp /2) \xi_{\nu_e}  <  0,
	\label{eq:7}
\end{equation}
which is a first order effect of $\xi_{\nu_e}$.  Having calculated these 
two effects (6) and 
(7), we can summarize qualitatively that the net \he4 abundance decreases 
if electron-neutrino 
are partially degenerate with positive $\xi_{\nu_e}$.

\section{Numerical Result of Primordial Nucleosynthesis}

\begin{figure}[ht]
\begin{center}
\epsfxsize=8.7cm 
\epsfysize=10cm
\epsfbox{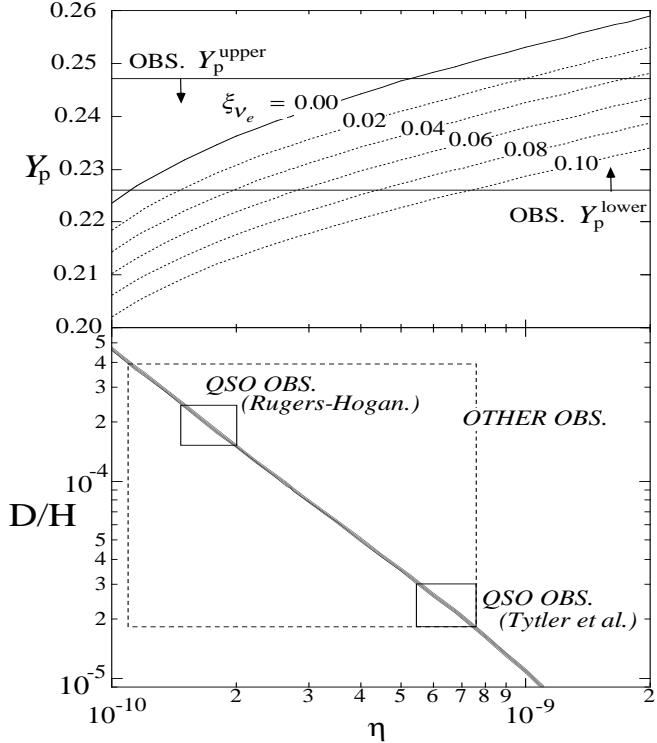}
\caption{$\yp$ and D/H versus $\eta$ for various degeneracy parameters 
of electron-neutrino 
$\xi_{\nu_e}$ in the homogeneous Big-Bang model. Observed abundance 
constraints on D/H are from Rugers-Hogan~\protect\cite{rugers96}, 
Tytler et al.~\protect\cite{tytler96}, and references therein.}
\label{fig1}
\end{center}
\end{figure}

Important result of the last section is that \he4 abundance is 
reduced for partial electron-neutrino degeneracy with arbitrary $\omegab$. 
Since \he4 abundance increases 
monotonically with increasing $\omegab$ as shown in Figure~\ref{fig1}, allowed 
$\omegab$ can be 
larger for satisfying the same \he4 abundance constraint. Let us quantify 
this by fulfilling the 
full network calculation of primordial nucleosynthesis.

\subsection{Homogeneous Model}

Shown in Figure~\ref{fig1} is the comparison between the observed abundance 
constraints on \he4 and D and the calculated curve in the homogeneous 
Big-Bang model as a function of $\eta$ 
for various $\xi_{\nu_e}$, where $\eta$ is the baryon-to-photon ratio, 
$\eta = n_B / n_\gamma$, and is linearly proportional to 
$\omegab$, $\omegab = 1.464 \times 10^8 \eta h^{-2}_{50}$. 
Calculated abundance curves with 
$\xi_{\nu_e}$ = 0 cannot find concordant $\omegab$ to satisfy both \he4 and D. 
Note that mass fraction of \he4, \yp, is displayed in linear scale, 
while the deuterium abundance relative to 
hydrogen, D/H = $n_{\rm D} / n_{\rm H}$, is shown in logarithmic scale. 
Although observed deuterium abundance from many different samples 
scatters largely, 
the constraint on \he4, 0.226 $\le \yp \le$ 0.247, is very accurate. 
This fact helps determine 
the most likely $\xi_{\nu_e}$.  The homogeneous model with $\xi_{\nu_e} 
\sim 0.05$ can best fit \he4 abundance as well as low deuterium abundance 
D/H $\sim 10^{-5}$.
\begin{figure}[ht]
\epsfxsize=16cm 
\epsfysize=7cm
\epsfbox{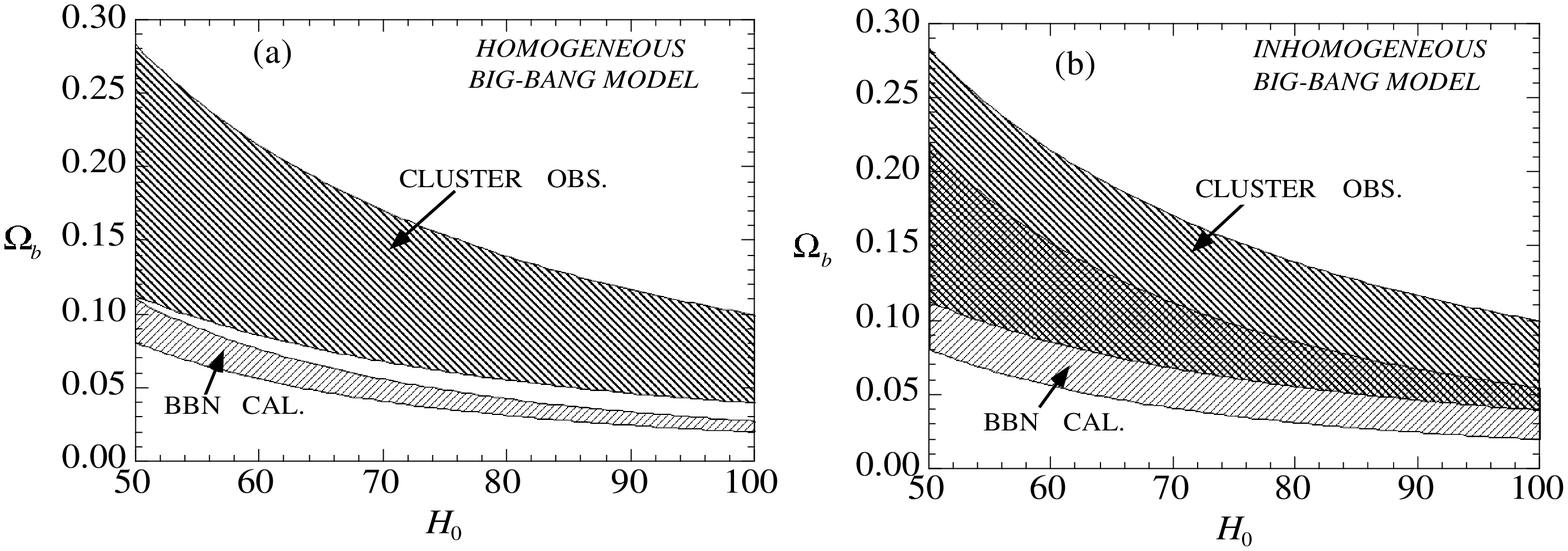}
\caption{Allowed $\omegab$ value versus $H_0$, inferred from 
astronomical observations of rich clusters of galaxies (CLUSTER 
OBS)~\cite{white93} 
and from the present theoretical calculations in (a) homogeneous and 
(b) inhomogeneous Big-Bang models (BBN CAL).}
\label{fig2}
\end{figure}

Unfortunately, however, the concordant $\Omega_b$ does not 
reach the observed value detected as hot X-ray gas, 0.1 
$\le \Omega_b h_{50}^{3/2} \le$ 0.3, in rich clusters of 
galaxies.  Figure~\ref{fig2}(a) shows dependence of both theoretically 
and observationally inferred $\Omega_b$ on the Hubble constant. 
The homogeneous model fails in 
explaining observed large $\Omega_b$ value in reasonable range of the Hubble 
constant 0.5 $\le h_{50} \le$ 1.0.

\subsection{Inhomogeneous Model}

It was found in our recent studies~\cite{orito97,mathews96} that the 
inhomogeneous Big-Bang model, which 
assumes non-degenerate neutrino, allows larger $\Omega_b$. 
If the inhomogeneous baryon-density distribution is created in some 
elementary process like cosmic QCD phase transition 
and others and if the scale of density inhomogeneities is 
comparable to the diffusion length of 
neutrons at the epoch of primordial nucleosynthesis, then 
$\Omega_b$ dependence of primordial nucleosynthesis is very different 
from the homogeneous model~\cite{appligate85}--\cite{kajino90}.  
Neutrons can easily diffuse out of the fluctuations for charge 
neutrality, while charged nuclei stay almost 
inside the fluctuations, thus forming specific density 
distributions. In high-density proton 
rich region is the neutron-to-proton ratio even smaller due to 
prominent neutron 
diffusion. \he4 production is suppressed here. On the other hand, 
\he4 production is regulated by small amount of protons. 
Thus the net \he4 production is reduced 
in the inhomogeneous model, though both diffusion and 
back-diffusion of all nuclides including neutrons make the process 
more complicated~\cite{orito97,mathews96}.

The diffusion of all nuclides including neutrons starts 
operating when electrons and positrons 
pair annihilate shortly after the weak decoupling and interconversion 
between protons and neutrons stops.  Degeneracy of electron-neutrino 
determines the neutron-to-proton ratio at 
this epoch before efficient diffusion starts. 
Therefore, nucleosynthesis is affected by the 
neutrino degeneracy in the inhomogeneous Big-Bang model as well.  
\begin{figure}[ht]
\begin{center}
\epsfxsize=12cm 
\epsfbox{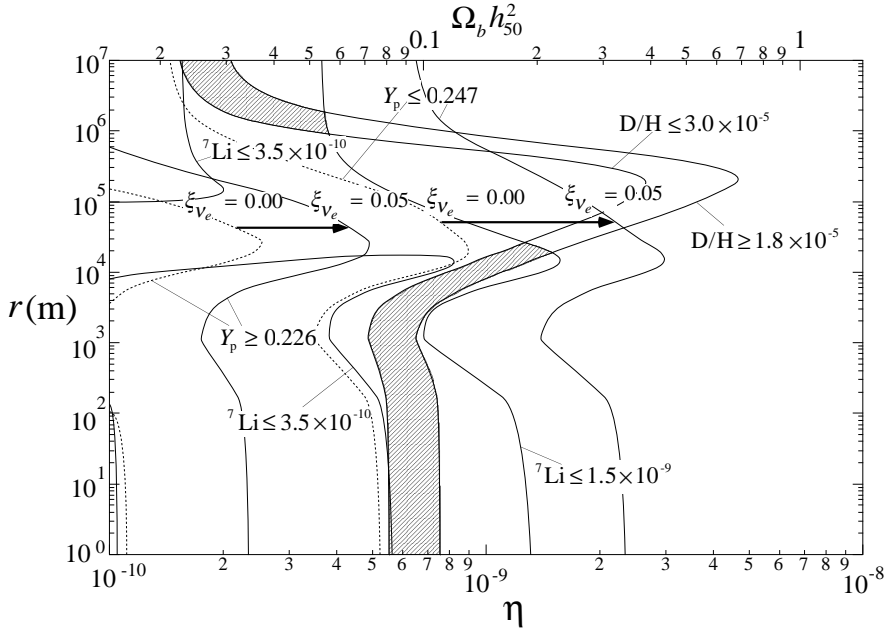}
\caption{Contours of allowed values for $\eta$ or $(\omegab\,h_{50}^2$) 
and fluctuation separation radius $r$ (at $T=1 \mev$). This 
calculation is based upon baryon density fluctuations represented by 
cylindrical shells (see ref.~\protect\cite{orito97} for details). The hatched 
region satisfies the light-element abundance constraints, as indicated, for 
degeneracy parameter of electron-neutrino $\xi_{\nu_e} = 0.05$.}
\label{fig4}
\end{center}
\end{figure}

The primordial nucleosynthesis calculated in the inhomogeneous model 
with $\xi_{\nu_e}$ 
= 0.0 and 0.05 are displayed in Figure~\ref{fig4} where contour map of each 
light element abundance is 
shown in the $r-\Omega_b (r-\eta)$ plane, where $r$ is the mean 
separation radius 
between fluctuating high-density regions at the $T = 1 \mev$ epoch.  
For the same mechanism as in the homogeneous 
model, \he4 abundance pattern changes drastically, while the other 
elements do not change very much.  Only when $\xi_{\nu_e}$ = 0.05 
is adopted, $\Omega_b h_{50}^2$ as large 
as 0.22 is allowed for $r \sim 10^4$ meter (at $T = 1 \mev$), 
even satisfying the light element 
abundance constraints whose detail is 
reported elsewhere. This allowed value of $\omegab$ is now in reasonable 
agreement with astronomical observation of hot X-ray gas, 0.1 
$\le \Omega_b h_{50}^{3/2} \le$ 0.3, as clearly shown in 
Figure~\ref{fig2}(b).

\section{Conclusion}

We studied the effects of lepton asymmetry of partially degenerate 
electron-neutrino on the 
primordial nucleosynthesis. Homogeneous Big-Bang model with 
neutrino degeneracy parameter $\xi_{\nu_e}$ = 0.05 can recover 
the concordance between 
\he4 and low deuterium abundance which was found in 
some Lyman-$\alpha$ 
clouds along the line of sight to high red-shift quasars, but 
the resultant $\Omega_b$ is less 
than detected in rich clusters 0.1 $\le \Omega_b h_{50}^{3/2} \le$ 0.3. 
It was found that the 
inhomogeneous Big-Bang model with the same degeneracy parameter can predict 
$\Omega_b$ as large as 0.22, which is in reasonable agreement 
with observation.  This 
degeneracy parameter corresponds to a small chemical potential 
of electron-neutrino of order 
10$^{-5}$ eV.  It is desirable to detect the asymmetry of 
background neutrinos.

\end{document}